# Optically Reconfigurable Photonic Devices

Qian Wang[1,2], Edward T. F. Rogers[1,3], Behrad Gholipour[1,4], Chih-Ming Wang[1], Guanghui Yuan[4], Jinghua Teng[2], and Nikolay I. Zheludev[1,4*]

[1]Optoelectronics Research Centre and Centre for Photonic Metamaterials, University of Southampton, Highfield, Southampton, SO17 1BJ, UK
[2]Institute of Materials Research and Engineering, Agency for Science, Technology and Research (A*STAR), 3 Research Link, Singapore 117602, Singapore
[3]Instiute for Life Sciences, University of Southampton, Highfield, Southampton, SO17 1BJ, UK
[4]Centre for Disruptive Photonic Technologies, Nanyang Technological University, Singapore 637371, Singapore
[*]niz@orc.soton.ac.uk

**Optoelectronic components with adjustable parameters, from variable-focal–length lenses to spectral filters that can change functionality upon stimulation, have enormous technological importance. Tuning of such components is conventionally achieved by either micro- or nano-mechanical actuation of their constitutive parts, stretching or application of thermal stimuli. Here we report a new dielectric metasurface platform for reconfigurable optical components that are created with light in a non-volatile and reversible fashion. Such components are written, erased and re-written as two-dimensional binary or grey-scale patterns into a nanoscale film of phase change material by inducing a refractive-index-changing phase-transition with tailored trains of femtosecond pulses. We combine germanium-antimony-tellurium-based films optimized for high-optical-contrast ovonic switching with a sub-wavelength-resolution optical writing process to demonstrate technologically relevant devices: visible-range reconfigurable bi-chromatic and multi-focus Fresnel zone-plates, a super-oscillatory lens with sub-wavelength focus, a grey-scale hologram and a dielectric metamaterial with on-demand reflection and transmission resonances.**

A metasurface made of carefully designed discrete metallic or dielectric elements (meta-molecules) can exhibit exceptional abilities for directing the flow of electromagnetic radiation across the entire electromagnetic spectrum with similar capabilities to planar holograms in optics [1, 2, 3, 4, 5, 6, 7]. As substantial efforts are now focused on developing metamaterials with switchable [8, 9] and reconfigurable metamolecules [10] driven by thermal [11, 12], electrostatic [13] and magnetic forces [14, 15], stretching [16] and light [17], we are witnessing the emergence of concepts of randomly accessible reconfigurable metamaterials in the microwave [18, 19] and optical regions of the spectrum [20, 21] thus making reconfigurable photonic devices controllable by external signals a realistic possibility. Here we introduce and demonstrate dynamic and non-volatile photonic components written into a dielectric metasurface that can be randomly and reversibly reconfigured with light. The randomly reconfigurable metasurface uses phase-change material and is written, erased and re-written as a two-dimensional binary or grey-scale pattern into a nanoscale thin film by inducing a refractive-index-changing phase-transition with tailored trains of femtosecond pulses.

We use phase-change medium the chalcogenide compound $Ge_2Sb_2Te_5$ (GST) which is widely exploited in rewritable optical disk storage technology and non-volatile electronic memories due to its good thermal stability, high switching speed and large number of achievable rewriting cycles [22, 23, 24]. When annealed to a temperature between the glass-transition and the melting point, the GST transforms from an amorphous state into a metastable cubic crystalline state, while a short high-density laser pulse melts and quickly quenches the material back to its amorphous phase with a pronounced contrast of dielectric properties observed between the two phases. Although nanosecond and microsecond laser pulses, which provide robust switching between amorphous and crystalline phases, are conventionally used in optical data storage technology, it was recently shown that femtosecond laser pulses can induce gradual (ovonic) switching [25, 26]. In the ovonic regime, metastable semi-crystallized states are created by carefully controlling the energy and the number of stimulating optical pulses. Here we use this regime with a high-repetition-rate femtosecond laser generating trains of a controlled number of 85 fs pulses at a wavelength of 730 nm to ovonically switch a thin film of GST. We demonstrate that light-induced phase-transition can be achieved in an extremely small volume of GST, down to 0.02 $\mu m^3$. Moreover, we show that the inverse phase-transition (from the crystalline to the amorphous state) can be achieved with the same femtosecond laser by modifying the excitation conditions. These techniques give us a toolkit for writing high-finesse functional patterns into the phase change films and erasing them as required.

## Results

**Writing of reconfigurable photonic devices in a phase change film**

To implement the technology we developed an apparatus comprising of sub-wavelength resolution optical-pattern generator/imaging system with writing controlled by a spatial light modulator and femtosecond laser source (see supplementary section S1 for details). The system uses a pulse-picker to control the intensity and duration of femtosecond pulse trains. We achieved resolution of 0.59 μm in the writing channel using lens with NA=0.8 (Fig. 1). To control the degree of ovonic change we formed trains of optical pulses of different duration, typically consisting of a few tens of 85 fs pulses. Although the femtosecond laser source provided optical pulses at 80 MHz repetition rate, we reduced this down to 1 MHz using the pulse picker. We used a Ti:Sapphire laser with single femtosecond pulse fluence up to 600 mJ/$cm^2$ at the wavelength of 730 nm. The written patterns were observed and recorded with LED illumination at 633 nm using a lens of NA= 0.7.

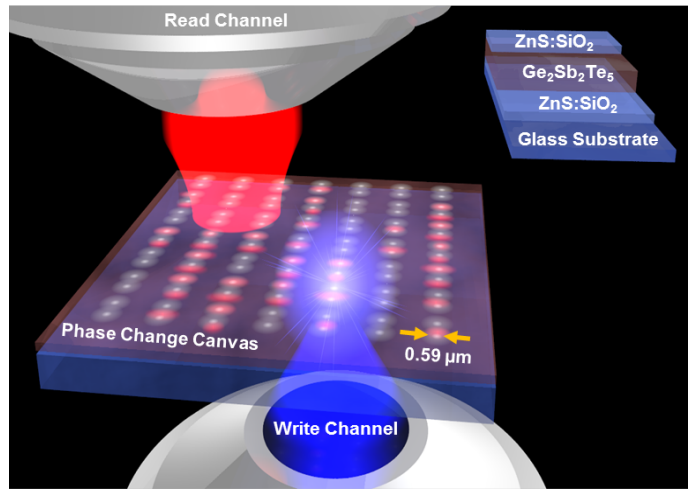

**Figure 1 Writing of reconfigurable photonic devices in a phase change film (artistic impression).** Various optical components, including lenses, diffractive elements and resonant metamaterials can be written with high accuracy in the chalcogenide glass phase-change film by trains of femtosecond pulses ("write" channel). Pulses from a Ti: Sapphire laser are focused and repositioned across the surface of the film by a computer-controlled spatial light modulator and electro-optical pulse picker. Optical excitation permanently changes the complex refractive index of the film by converting continuously from the amorphous to crystalline state in allowing films with complex refractive-index profiles to be written. The written pattern can also be erased by the same laser using different illumination conditions. The results are observed through the "read" channel.

All our experiments were performed with 70 nm thick GST films sputtered on a glass substrate covered with an intermediate $ZnS-SiO_2$ film. After deposition, the GST film was covered with another protective layer of $ZnS-SiO_2$.

Using the apparatus described above, we demonstrate several optical components written on demand in the GST film. A Fresnel zone-plate is a planar focusing device that can be used in the confined environment of integrated optical circuits and imaging devices. Here we show that good quality binary Fresnel zone-plates can be optically written, erased and re-written. We also demonstrate writing wavelength multiplexing focusing devices and chromatically corrected focusing zone-plates. The high finesse of our techniques also allows the demonstration of a binary optically-written super-oscillatory lens focusing light beyond the diffraction limit. The ability to control the degree of crystallization in the GST film has allowed us to create an 8-level grey-scale hologram producing a multi-spot pattern. And finally, we used optical writing to create a planar dielectric metamaterials with optical resonances in the near infrared part of the spectrum.

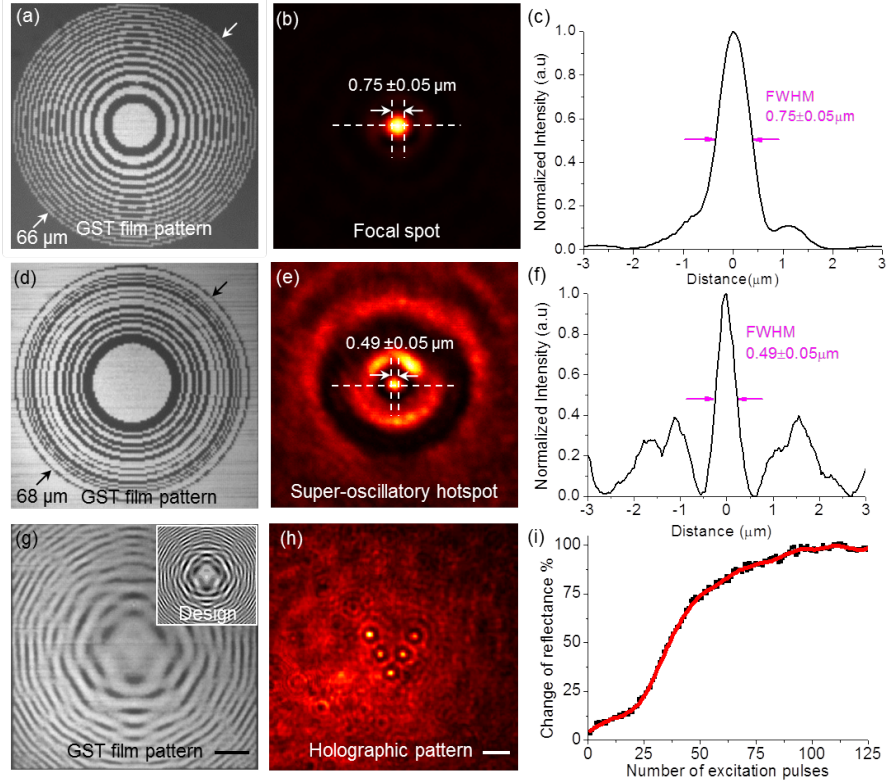

**Figure 2 Binary and grey-scale devices optically written in the phase-change film. a, Fresnel zone-plate pattern** imaged at λ= 633 nm; **b,** Microscope image of the optical hotspot as focused by the Fresnel zone-plate (a); **c,** Intensity cross-section of the hot-spot (b); **d, Binary Super-oscillatory Lens pattern** imaged at λ= 633 nm; **e,** Microscope image of the optical hotspot as focused by the Binary Super-oscillatory Lens (d) formed in transmission mode at 43.8 μm from the lens; **f**, Intensity cross-section of the hot-spot (e); **g**, Image of an 8-level **grey-scale hologram** designed to generate a V-shaped five-spot pattern. Inset is the computer generated grey-scale hologram with 121×121 pixels. The scale bar is 10 μm; **h**, Microscope image of the generated five spot pattern in transmission mode at 100 μm away from the sample surface, λ= 730 nm. The scale bar is 5 μm; **i**, Continuous, ovonic change of reflectance, as given by $\Delta R=(R_c-R_a)/R_a$, of the partially crystallized chalcogenide glass film as a function of the number of femtosecond pulses exciting the film. Here $R_a$ is reflectance of the amorphous film, and $R_c$ is reflectance of the partially crystallized mark. The single pulse energy is 0.39 nJ, corresponding to fluence of approximately 140 mJ/cm$^2$.

## Binary and grey-scale photonic devices

A Fresnel zone-plate consists of a set of alternately opaque and transparent radially symmetric rings used to focus light by diffraction. The zones are designed to contribute equal intensity to the focal spot. Figure 2a shows an optical image of the Fresnel zone-plate consisting of 27 Fresnel zones that has the diameter of 66 μm and focal length $f$=50 μm. The pattern of the plate is pixelated to 111×111 pixels and then transferred to the phase change film using trains of 80 pulses with each pulse delivering 0.39 nJ per point. The focusing performance of Fresnel zone-plate is characterized under the normal illumination of laser beam at λ=730 nm by a 100× lens with NA=0.9. The wave plate produces a clear, well defined focal spot of $r$=0.75±0.05 μm (FWHM) in Fig. 2b-c. This focal spot size is comparable with the estimation of hot spot that can be achieved by an ideal Fresnel zone-plate lens, $r_{FZP}$=1.22×Δr ~ 0.72 μm, where Δr is the smallest zone width, determined in this case by the mark size (0.59 μm) of our pattern, which agrees with the theoretical calculation in supplementary section S2. The slightly bigger hotspot in the experiment is probably due to the pixellation approximation and limited fabrication precision.

We also successfully created a super-oscillatory binary lens that requires considerable fabrication finesse to perform correctly. In the past such lenses were only fabricated by e-beam lithography and focused ion beam milling [27, 28]. The super-oscillatory lens provides tailored interference of propagating waves to create a sub-diffraction hotspot in the optical far-field. The lens, designed by the particle swarm optimization algorithm [28, 29], resembles a Fresnel zone-plate, but consists of 21 concentric rings of irregular width and diameter. The ring pattern with overall size of 68 μm is pixelated into 114×114 laser-written marks fabricated under same writing condition as Fresnel zone-plate (Fig. 2d). When illuminated at λ=730 nm, the super-oscillatory lens generates a central hotspot of $r$=0.49±0.05 μm (FWHM) at a distance of 43.8 μm from the mask characterized by a 150× lens with NA=0.95, as shown in Fig. 2e-f. It is comparable with the theoretically calculated value of 0.46 μm for this mask. The slight difference between the predicted and experimentally achieved hotspot width is due to pixellation of the mask and fabrication tolerances. The hotspot is beyond the diffraction-limited hotspot of 0.6 μm that could have been generated by a conventional lens with the same diameter and focal distance. Simulation results are provided in supplementary section S2.

Fresnel lenses and super-oscillatory device described above are formed from binary patterns (rings) that use only two phases of the GST film. However, thanks to the ovonic nature of the phase change in GST we can also write grey-scale patterns into a nanoscale film by continuously controlling the degree of phase transition. This allows us to induce gradual grey-scale variations of the refractive index across the film by controlling the degree of crystallization on each mark in the GST film through applying controlled number of laser writing pulses from mark to mark. A flat hologram can be created this way. We tested this by generating an 8-level grey-scale hologram that is designed to operate at 730 nm wavelength, generating V-shaped pattern of hot-spots at a distance of 100 μm from the mask (see Figure 2g).

To design the grey-scale hologram pattern we reversed the holographic image reconstruction process. We simulated five point sources located at a distance of 100 μm from the GST film with a plane wave reference beam. The required grey-scale hologram design of 71 μm × 71 μm in overall size was generated from the interference pattern created by these sources on the film. It was then pixelated to 8 grey-scale levels and used as the fabrication design. To write a mask with different grey levels, we use a variable number of laser pulses at each point, from 0 to 80, as calculated from Fig. 2i. The image stored in the hologram pattern was then reconstructed by illuminating it with a plane wave at λ=730 μm. In spite of the limited size of the hologram and pixellation, we observed a clear V-shaped five-spot pattern at a distance of 100 μm, as presented in Fig. 2h. The blurred background is partially due to light absorption in GST recording medium, which is not accounted for in the design of hologram pattern, but reduces the diffraction efficiency and diminishes the signal-to-noise ratio. Nevertheless, the ability to create a hologram by continuously controlling the phase change in the GST film is clearly demonstrated. Simulations of the reflection, transmission and absorption of the GST media are provided in supplementary section S3.

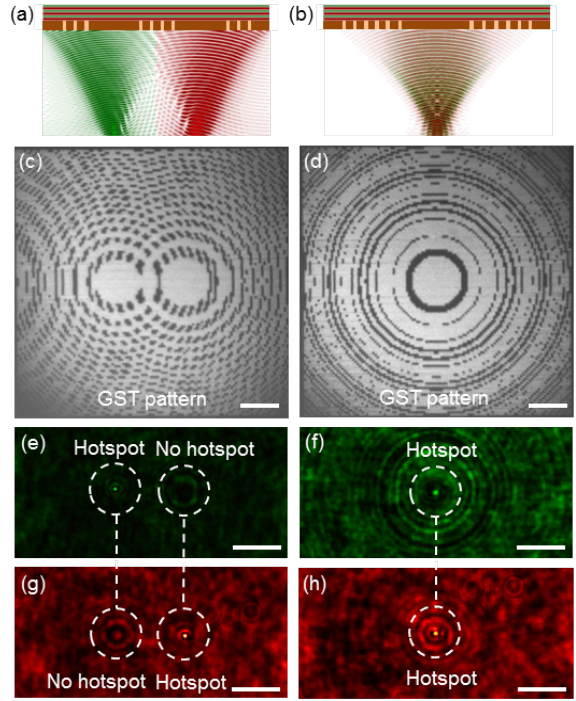

**Figure 3 Writing planar wavelength multiplexing focusing devices. a,** A lens focusing two different wavelengths to spatially separated foci on the focal plane (deliberate transverse chromatic aberration); **b**, A lens focusing two different optical wavelengths in the same focus (corrected chromatic aberration); **c,d** Optical images of the lens patterns in GST film with deliberate transverse chromatic aberration (c) and corrected chromatic aberration (d). Each pattern is composed of 121×121 pixels; **e,f**, focal spots of lenses (c) and (d) at $\lambda = 730$ nm; **g,h**, focal spots of the same lenses at $\lambda = 900$ nm. Note that pattern (c) focuses light of different wavelength in different spots, while pattern (d) focuses light of both wavelengths in the same spot position. The scale bar is 10 μm.

**Planar wavelength multiplexing focusing devices**

We have also demonstrated more sophisticated chromatically selective and chromatically corrected lenses presented on Fig 3a-b. The lens with deliberate transverse chromatic aberration is a double-foci lens focusing light with different wavelengths in distinctly different focal spots (Fig. 3a). Such compact planar lenses may be useful, for instance, in on-chip sensing and for signal multiplexing. In contrast, the chromatically corrected lens focuses light with two different wavelengths into a single hotspot (Fig. 3b), which may be useful in imaging and broadband light-detection applications.

The lens with deliberate chromatic separation consists of two slightly different, mutually displaced Fresnel zone-plate patterns (Fig. 3c). When illuminated at the wavelengths of 730 nm and 900 nm, it generates foci at the same focal distance of 50 μm but spatially separated by 17.8 μm (Fig. 3e, g). The chromatically corrected lens consists of two concentric Fresnel zone-plate patterns (Fig. 3d) designed to create foci for 730 nm and 900 nm wavelengths at the same focal distance of 50 μm (Fig. 3f, h). Both types of chromatically selective and chromatically corrected lenses show satisfactory performance and close to the theoretical hot-spot sizes, comparable with the simulation results presented in supplementary information, section S4.

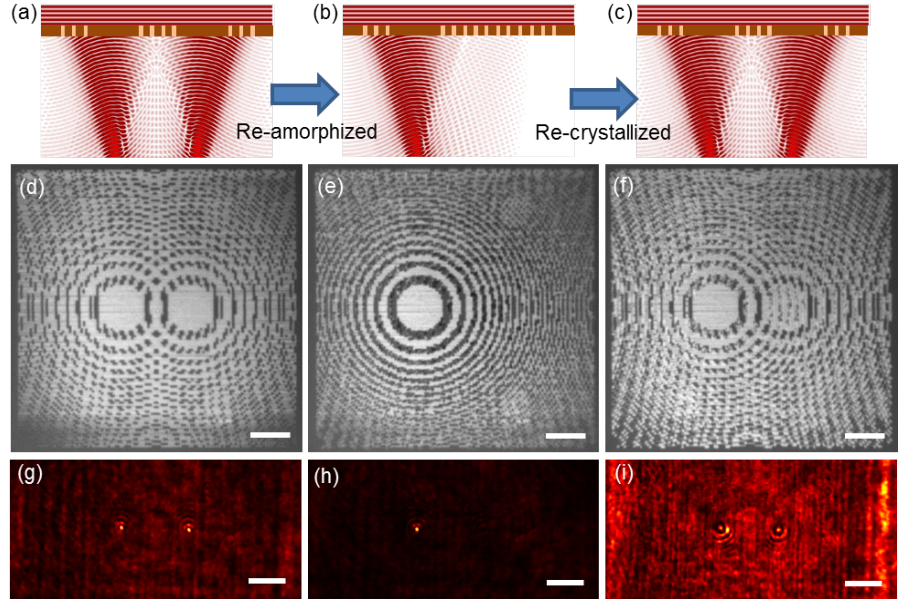

**Figure 4 Dynamically optically reconfigurable zone-plate device. a**, Two superimposed Fresnel zone patterns focusing plane wave into two different foci; **b,c,** one of the Fresnel zone patterns is erased (b) and then restored again (c); **d,** superimposed Fresnel zone patterns imaged at λ= 633nm as they are first written; **e,** the second Fresnel zone pattern is erased; **f,** both patterns are restored; **g-i,** transmission focal spots as generated by patterns (d-f) at λ= 730nm. The scale bar is 10 μm.

## Dynamically optically reconfigurable zone-plate device

The GST film can sustain a large number of transition cycles between amorphous and crystalline states, which is a major advantage for reconfigurable photonic devices. This allows different functionalities to be written into the same optical materials without changing the structure of the optical system. We demonstrated the write-erase-write reconfiguration cycle with a double Fresnel zone-plate pattern (Fig. 4a-c). We first wrote a double Fresnel zone-plate pattern of 121×121 pixels in size with 0.39 nJ pulses. Upon illumination with a plane wave at λ= 730 nm it generated two foci imaged by 50× lens with NA=0.75, as shown in Fig. 4d, g. We then erased one of the Fresnel zone-plates by re-amorphisation of its pattern pixel by pixel with single pulse of 445 mJ/cm$^2$ fluence (Fig. 4e, h). Finally, we demonstrated that the erased Fresnel zone-plate pattern can be restored by writing it again with 0.39 nJ pulses (Fig. 4f, i). We observed that the as-deposited films require larger energy (approximately 20%) to crystallize than that is required in subsequent cycles. The restored, re-crystallized pixels look slightly brighter than those crystallized in the first writing. Here we also note that, instead of point-to-point erasing, if necessary, the pattern can be erased in one step by using a single high energy pulse with large hot-spot [25], dramatically simplifying and speeding up the process.

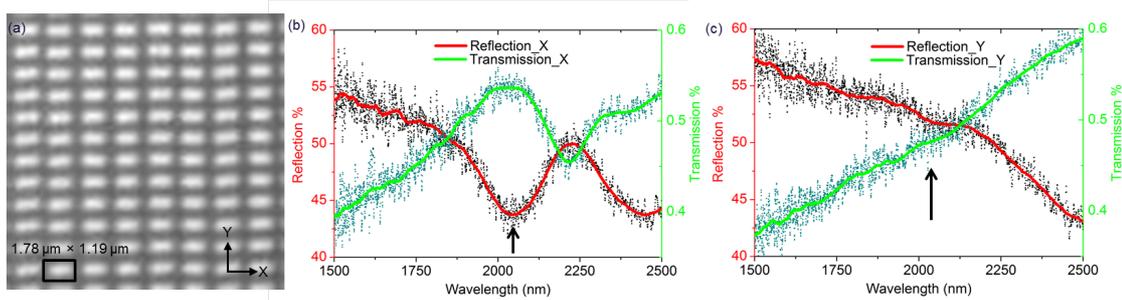

**Fig 5 Writing a dielectric metamaterial. a,** Reflection image of the dielectric metamaterial written into the GST phase change film; The 1.78 μm × 1.19 μm unit cell of the pattern consists of two phase-change marks; **b,c** Reflection (red line) and transmission (green line) spectra of the metamaterial for light polarized along horizontal direction (b) vertical direction (c), as indicated in (a).

**A dielectric metamaterial**

One of the most useful properties of GST materials is the huge difference of refractive index between amorphous and crystallized states in near-infrared region [30]. This allows on-demand writing of dielectric metamaterial patterns with resonances at near-infrared frequencies where the available spatial resolution is sufficient to write non-diffracting two-dimensional arrays of sub-wavelength meta-molecules. Moreover, in the spectral range between 2 and 2.2 microns amorphous GST has absorption less than 0.3 $\mu m^{-1}$ which is important to achieve high-quality resonances [31]. We demonstrate a dipolar dielectric metamaterial, a two-dimensional array of rectangular crystalline inclusions in the amorphous film (seen in Fig. 5a). Each inclusion was composed of two partially-overlapping laser-written crystalline marks, as shown in Fig. 5a. Reflection and transmission measurements using a Fourier transform infrared spectrometer show the resonant transmission peak and reflection dip at 2040 nm for light polarized along the long axis of the inclusion (Fig. 5b) and no resonance features for the orthogonal polarization (Fig. 5c). We emphasize that these resonances are of a true metamaterial nature: at wavelength longer than 1.78 μm the structure does not diffract and may be considered an "effective medium". These experimental results agree well with the full 3D Maxwell simulation results presented in supplementary material section S5, which show similar calculations of transmission and reflection from the metamaterial array. We note that in the simulations, the resonance is at a slightly different wavelength of 2460 nm. We argue that this discrepancy could be explained by small imperfections in the written pattern, inaccuracies in the permittivity data used in the simulation and an incomplete phase transition from amorphous to crystalline in the experimental case that would lead to a smaller refractive index value than used in the calculations for crystallized GST.

## Discussions

In conclusion, we have reported a new and versatile platform for creating dynamically reconfigurable optical devices that can be reconfigured with light and illustrated this technology with a number of demonstrations including focusing devices and resonant metamaterials. This femtosecond-laser-controlled, reversible, multi-level phase-switching is a robust and flexible technology that builds on well-established and reliable optical phase change disk-technology. In addition to creating dynamically reconfigurable optical devices, it allows applications in high-density optical data storage and image processing.

# Methods

### Sample preparation
The phase change stack was deposited on pre-cleaned glass microscope slides by RF sputtering in a Nano 38 system (Kurt J. Lesker, USA) from 2 inch wide targets (Testbourne, UK). The deposition starts with a layer of ZnS/SiO$_2$ (2:8) of 70 nm thick followed by the deposition of the phase change layer Ge$_2$Sb$_2$Te$_5$ at a power of 45 W. A base pressure of $5 \times 10^{-5}$ mbar is achieved prior to deposition. The sputtering gas used is high-purity argon. A flow rate of 70 ccpm is used to strike a plasma on all targets, whilst 37 ccpm of the same gas is used to maintain the plasma during deposition, in the case of GST. This is followed by a 70 nm deposition of the capping layer of ZnS/SiO$_2$ (2:8) without breaking vacuum at an RF power 45 W in the presence of 15 ccpm Ar during deposition. The substrate is held on a rotating platen with a target–substrate distance of ~150 mm which gives deposits of low stress. The substrates, initially at room temperature, are heated by <10 K during deposition.

## Acknowledgements

The authors thank A. Karvounis for assistance in computing spectra of dielectric metamaterial and D. Hewak for useful discussion. This study was supported by the Engineering and Physical Sciences Research Council UK (grant EP/G060363/1), the Singapore Ministry of Education (Grant MOE2011-T3-1-005) and the Agency for Science, Technology and Research (A*STAR) of Singapore (Grants 122-360-0007, 122-360-0009) and the University of Southampton Enterprise Fund. Q. Wang acknowledges the fellowship support from the A*STAR.

## Author contributions

N. I. Z. conceived the idea of optical reconfigurable photonics devices; Q. W. built up the experimental setup and carried out the experiments; Q. W. and E. T. F. R. designed the experimental apparatus and carried out data analysis; B. G. prepared the experimental samples; C. M. W. designed the hologram pattern; G. H. Y. designed the super-oscillatory lens and performed angular spectrum simulations; Q. W., N. I. Z. and J. H. T. co-wrote the paper. All coauthors discussed the results and cross-edited the manuscript. N. I. Z. and J. H. T. supervised and coordinated all the work.

## Additional information

Supplementary Information is available. Correspondence and requests for materials should be addressed to Q. Wang. The data from this paper can be obtained from the University of Southampton ePrints research repository: http://dx.doi.org/10.5258/SOTON/377671

## Competing financial interests

The authors declare no competing financial interests.